\documentclass[apj]{emulateapj}
\slugcomment{accepted by ApJ}

\shorttitle{LARGE-SEPARATION LENSES}
\shortauthors{OGURI \& KEETON}

\begin{document}

%%%%%%%%%%%%%%%%%%%%%%%%%%%%%%%%%%%%%%%%%%%%%%%%%%%%%%%%%%%%%%%%%%%%%%
%%%%%%%%%%%%%%%%%%%%%%%%%%%%%%%%%%%%%%%%%%%%%%%%%%%%%%%%%%%%%%%%%%%%%%
\title{Effects of Triaxiality on the Statistics of Large-Separation
Gravitational Lenses} 
%%%%%%%%%%%%%%%%%%%%%%%%%%%%%%%%%%%%%%%%%%%%%%%%%%%%%%%%%%%%%%%%%%%%%%
%%%%%%%%%%%%%%%%%%%%%%%%%%%%%%%%%%%%%%%%%%%%%%%%%%%%%%%%%%%%%%%%%%%%%%

%%%%%%%%%%%%%%%%%%%%%%%%%%%%%%%%%%%%%%%%%%%%%%%%%%%%%%%%%%%%%%%%%%%%%%
\author{Masamune Oguri\altaffilmark{1} and Charles R. Keeton\altaffilmark{2,3}}
\altaffiltext{1}{Department of Physics, University of Tokyo, Hongo 7-3-1,
Bunkyo-ku, Tokyo 113-0033, Japan.}
\altaffiltext{2}{Astronomy and Astrophysics Department, University of
Chicago, 5640 South Ellis Avenue, Chicago, IL 60637 USA.}
\altaffiltext{3}{Hubble Fellow}
\email{oguri@utap.phys.s.u-tokyo.ac.jp (MO), \\
ckeeton@oddjob.uchicago.edu (CRK)}
%%%%%%%%%%%%%%%%%%%%%%%%%%%%%%%%%%%%%%%%%%%%%%%%%%%%%%%%%%%%%%%%%%%%%%

%\received{}
%\accepted{}

\begin{abstract}
We study the statistics of large-separation gravitational lens
systems produced by non-spherical halos in the Cold Dark Matter
(CDM) model.  Specifically, we examine how the triaxiality of CDM
halos affects the overall lensing probabilities and the relative
numbers of different image configurations (double, quadruple,
and naked cusp lenses).  We find that triaxiality significantly
enhances lensing probabilities by a factor of $\sim$2--4, so it
cannot be ignored.  If CDM halos have central density slopes
$\alpha \lesssim 1.5$, we predict that a significant fraction
($\gtrsim$20\%) of large-separation lenses should have naked cusp
image configurations; this contrasts with lensing by isothermal
($\alpha \approx 2$) galaxies where naked cusp configurations are
rare.  The image multiplicities depend strongly on the inner
density slope $\alpha$: for $\alpha=1$, the naked cusp fraction
is $\gtrsim$60\%; while for $\alpha=1.5$, quadruple lenses are
actually the most probable.  Thus, the image multiplicities in
large-separation lenses offer a simple new probe of the inner
density profiles of dark matter halos.  We also compute the
expected probabilities and image multiplicities for lensed
quasars in the Sloan Digital Sky Survey, and argue that the recent
discovery of the large-separation quadruple lens SDSS J1004+4112
is consistent with expectations for CDM.
\end{abstract}

%%%%%%%%%%%%%%%%%%%%%%%%%%%%%%%%%%%%%%%%%%%%%%%%%%%%%%%%%%%
%%%%%%%%%%%%%%%%%%%%%%%%%%%%%%%%%%%%%%%%%%%%%%%%%%%%%%%%%%%
\section{Introduction}
%%%%%%%%%%%%%%%%%%%%%%%%%%%%%%%%%%%%%%%%%%%%%%%%%%%%%%%%%%%
%%%%%%%%%%%%%%%%%%%%%%%%%%%%%%%%%%%%%%%%%%%%%%%%%%%%%%%%%%%

The Cold Dark Matter (CDM) model of structure formation naturally
predicts the existence of strong gravitational lens systems with
image separations of $\sim\!10''$ or even larger.  Observations
of massive clusters of galaxies have revealed many systems of
``giant arcs'' representing lensed images of background galaxies
\citep{lynds86,soucail87,luppino99,gladders03,zaritsky03}.
However, until recently all lensed quasars and radio sources had
image separations $<\!7''$ corresponding to lensing by galaxies,
despite some explicit searches for lenses with larger separations
\citep{phillips01,ofek01}.\footnote{\citet{miller04} recently
reported six pairs of quasars in the Two-Degree Field (2dF) Quasar
Redshift Survey that are candidate lenses with image separations on the
scale of an arcminute; but none of the candidates has been confirmed,
and theoretical arguments by \citet{oguri03a} indicate that it would
be quite surprising if any of the systems are lenses.}  Lensing of
quasars by clusters was finally observed with the recent discovery
and confirmation of SDSS J1004+4112, a quadruple lens with an
image separation of $14\farcs62$ found in the Sloan Digital Sky
Survey \citep{inada03,oguri04}.  This lens confirms an important
prediction of the CDM model; indeed, the lensing probability
inferred from the discovery is in agreement with reasonable values
of the cosmological parameters \citep{oguri04}.

The statistics of large-separation lenses can be used to place
constraints on the density profile of dark halos \citep*{maoz97,
keeton01c,keeton01a,wyithe01,takahashi01,sarbu01,li02,oguri02a,
oguri02b,huterer04,kuhlen04}, or determine the abundance of massive
dark halos \citep{narayan88,wambsganss95,kochanek95,nakamura97,
mortlock00,oguri03a,lopes04,chen04}.  The discovery of the first
such lens suggests that these statistics can be a practical tool
to study structure formation in the universe.  The statistics of
giant arcs are also known as a good probe of clusters
\citep*{bartelmann98,meneghetti01,molikawa01,oguri01,oguri03b,
wambsganss04,dalal04,maccio04}, and in fact lensed arcs and quasars
complement each other in several ways.  For instance, in lensed
quasar surveys one first identifies source quasars and then checks
whether they are lensed, while in searching for lensed arcs one
selects massive clusters and then searches for lensed arcs in
them.  In other words, surveys for arcs are biased toward high
mass concentrations, while lensed quasars probe random lines of
sight.  Clusters selected by the presence of lensed quasars could,
in principle, differ from those selected as having giant arcs.  In
addition, lensed quasars have three advantages over lensed arcs
in statistical studies.  First, quasars can be regarded as point
sources, while sources for arcs are galaxies whose intrinsic sizes
and shapes are important but unobservable.  Second, the number and
configuration of images in a quasar lens system is unambiguous.
Third, the redshift distribution of arc sources is poorly known
(and controversial; see \citealt{oguri03b}; \citealt{wambsganss04};
\citealt{dalal04}), 
while the redshift distribution of quasars is well known.

In all previous analytic work on the statistics of large-separation
lensed quasars, the lens objects were assumed to be spherical.
However, in the CDM model dark halos are not spherical at all but
triaxial \citep[e.g,][hereafter JS02]{jing02}.  It is already known
that triaxiality has a significant effect on the statistics of
lensed arcs, from both analytic \citep[][hereafter OLS03]{oguri03b}
and numerical (\citealt*{meneghetti03a}; \citealt{dalal04}) points
of view.  In the statistics of normal lensed quasars, triaxiality
(or ellipticity) has been thought to mainly affect the image
multiplicities, with only small changes to the total lensing
probability \citep*{kochanek96,keeton97,evans02,chae03,hk04}.
However, that conclusion is based on nearly-singular isothermal
lens models, and the situation may be quite different for the less
concentrated mass distributions of the massive halos that create
large-separation lenses.  Moreover, only triaxial modeling allows
us to study image multiplicities, and to consider whether it is
statistically natural that the first known large-separation lens
is a quadruple.

The structure of this paper is as follows.  In \S\ref{sec:theory}
we review the triaxial dark halo model and its lensing properties
(originally presented by JS02 and OLS03).  In \S\ref{sec:prob} we
show our general results, while in \S\ref{sec:sdss} we customize
our predictions to the SDSS quasar sample.  We summarize our
conclusions in \S\ref{sec:sum}.  Throughout the paper, we assume
a $\Lambda$-dominated cosmology with current matter density
$\Omega_M=0.3$, cosmological constant $\Omega_\Lambda=0.7$,
dimensionless Hubble constant $h=0.7$, and normalization of matter
density fluctuations $\sigma_8=0.9$.

%%%%%%%%%%%%%%%%%%%%%%%%%%%%%%%%%%%%%%%%%%%%%%%%%%%%%%%%%%%
%%%%%%%%%%%%%%%%%%%%%%%%%%%%%%%%%%%%%%%%%%%%%%%%%%%%%%%%%%%
\section{Formalism}\label{sec:theory}
%%%%%%%%%%%%%%%%%%%%%%%%%%%%%%%%%%%%%%%%%%%%%%%%%%%%%%%%%%%
%%%%%%%%%%%%%%%%%%%%%%%%%%%%%%%%%%%%%%%%%%%%%%%%%%%%%%%%%%%

%%%%%%%%%%%%%%%%%%%%%%%%%%%%%%%%%%%%%%%%%%%%%%%%%%%%%%%%%%%
\subsection{Lensing by triaxial dark halos}\label{sec:lensing}
%%%%%%%%%%%%%%%%%%%%%%%%%%%%%%%%%%%%%%%%%%%%%%%%%%%%%%%%%%%

In this section, we briefly summarize the lensing properties of the
triaxial model of dark halos proposed by JS02. For more details,
please refer to JS02 and OLS03.

First, we relate the principal coordinate system of the triaxial
dark halo $\vec{x} = (x,y,z)$ to the observer's coordinate system
$\vec{x'} = (x',y',z')$, where the $z'$-axis runs along the line
of sight to the observer.  In general, the coordinate
transformation is expressed as $\vec{x}=A\vec{x'}$ with
%%%%%%%%%%%%%%%%%%%%%%%%%%%
\begin{equation}
A\equiv\left(
\begin{array}{ccc}
 -\sin\phi & -\cos\phi\cos\theta & \cos\phi\sin\theta \\
 \cos\phi  & -\sin\phi\cos\theta & \sin\phi\sin\theta \\
 0         & \sin\theta          & \cos\theta\\
\end{array}
\right) .
\end{equation}
%%%%%%%%%%%%%%%%%%%%%%%%%%%

The density profiles of triaxial dark matter halos proposed by
JS02 \citep[also see][]{zhao96} is
%%%%%%%%%%%%%%%%%%%%%%%%%%%
\begin{equation}
 \rho(R)=\frac{\delta_{\rm ce}\rho_{\rm crit}(z)}
{(R/R_0)^\alpha(1+R/R_0)^{3-\alpha}},
\label{gnfw}
\end{equation}
%%%%%%%%%%%%%%%%%%%%%%%%%%%
where
%%%%%%%%%%%%%%%%%%%%%%%%%%%
\begin{equation}
 R^2\equiv c^2\left(\frac{x^2}{a^2}+\frac{y^2}{b^2}
+\frac{z^2}{c^2}\right)\;\;\;(a\leq b\leq c).
\label{rdef}
\end{equation}
%%%%%%%%%%%%%%%%%%%%%%%%%%%
Two models commonly discussed in the context of CDM simulations
are $\alpha=1$ and 1.5 \citep*{navarro97,fukushige97,moore99,jing00,
power03,fukushige04}, and we focus on these.  JS02 give fitting
formulas for the axis ratios $a/c$ and $a/b$ in the triaxial model,
and for the concentration parameter $c_e\equiv R_e/R_0$, where $R_e$
is defined such that the mean density within the ellipsoid of the
major axis radius $R_e$ is $\Delta_e\Omega(z)\rho_{\rm crit}(z)$
with $\Delta_e = 5\Delta_{\rm vir}\left(c^2/ab\right)^{0.75}$.

What matters for lensing is the projected surface mass density
in units of the critical density for lensing, or the convergence
$\kappa$, which can be expressed as (OLS03)
%%%%%%%%%%%%%%%%%%%%%%%%%%%
\begin{equation}
 \kappa=\frac{b_{\rm TNFW}}{2}\
   f_{\rm GNFW}\left(\frac{1}{R_0}\sqrt{\frac{(x')^2}{q_x^2}+
   \frac{(y')^2}{q_y^2}}\right),
\label{convergence}
\end{equation}
%%%%%%%%%%%%%%%%%%%%%%%%%%%
where $b_{\rm TNFW}$ is a dimensionless ``strength'' parameter
(defined in OLS03), $q \equiv q_y/q_x \le 1$ is the axis ratio of
the projected mass distribution, and
%%%%%%%%%%%%%%%%%%%%%%%%%%%
\begin{equation}
 f_{\rm GNFW}(r) \equiv
\int_{0}^{\infty}\frac{1}{\left(\sqrt{r^2+z^2}\right)^\alpha
\left(1+\sqrt{r^2+z^2}\right)^{3-\alpha}}\,dz.
\label{f_nfw}
\end{equation}
%%%%%%%%%%%%%%%%%%%%%%%%%%%
For $\alpha=1$, equation (\ref{f_nfw}) has an analytic expression
\citep{bartelmann96}.  For $\alpha=1.5$, we adopt a fitting
formula for equation (\ref{f_nfw}) presented by OLS03.  Note that the
convergence has elliptical symmetry, so the lensing deflection
and magnification can be computed with a set of 1-dimensional
integrals \citep{schramm90,keeton01b}.  Also note that if we work
in dimensionless units, scaling all lengths by $L_0 \equiv R_0 q_x$,
then the lensing properties of the dark matter halos depend only
on the parameters $\alpha$, $b_{\rm TNFW}$, and $q$.

%%%%%%%%%%%%%%%%%%%%%%%%%%%%%%%%%%%%%%%%%%%%%%%%%%%%%%%%%%%
\subsection{Cross sections and image separation distributions}\label{sec:cross}
%%%%%%%%%%%%%%%%%%%%%%%%%%%%%%%%%%%%%%%%%%%%%%%%%%%%%%%%%%%

%%%%%%%%%%%%%%%%%%%%%%%%%%%%%%%%%%%%%%%%%%%%%%%%%%%%%%%%%%%%%%%%%%%%%
\begin{figure*}
\epsscale{0.7}
\plotone{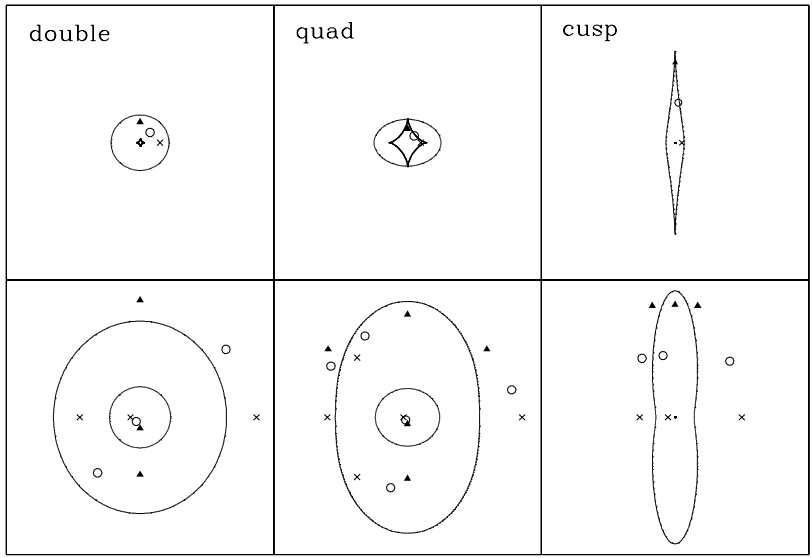}
\caption{Sample image configurations. The top panels show the source
planes, and the bottom panels show the corresponding image planes.
The solid lines indicate the caustics and critical curves.  We show
three sources (denoted by triangles, circles, crosses), and their
corresponding images.  From left to right, the lenses are doubles,
quadruples, and cusps. Specific values of $(b_{\rm TNFW}, q)$ for each
 example are $(2,0.95)$, $(2,0.75)$, and $(0.6,0.25)$ for doubles,
 quadruples, and cusps, respectively. Doubles and cusps are
 distinguished by the image parities: doubles have one positive-parity
 image and one negative-parity image, plus a central double-negative
 image that is usually too faint to be observed; while cusps have two
 positive-parity images and one negative parity image, all of comparable
 brightnesses.
\label{fig:multi}} 
\end{figure*}
%%%%%%%%%%%%%%%%%%%%%%%%%%%%%%%%%%%%%%%%%%%%%%%%%%%%%%%%%%%%%%%%%%%%%

\begin{figure*}
\epsscale{0.7}
\plotone{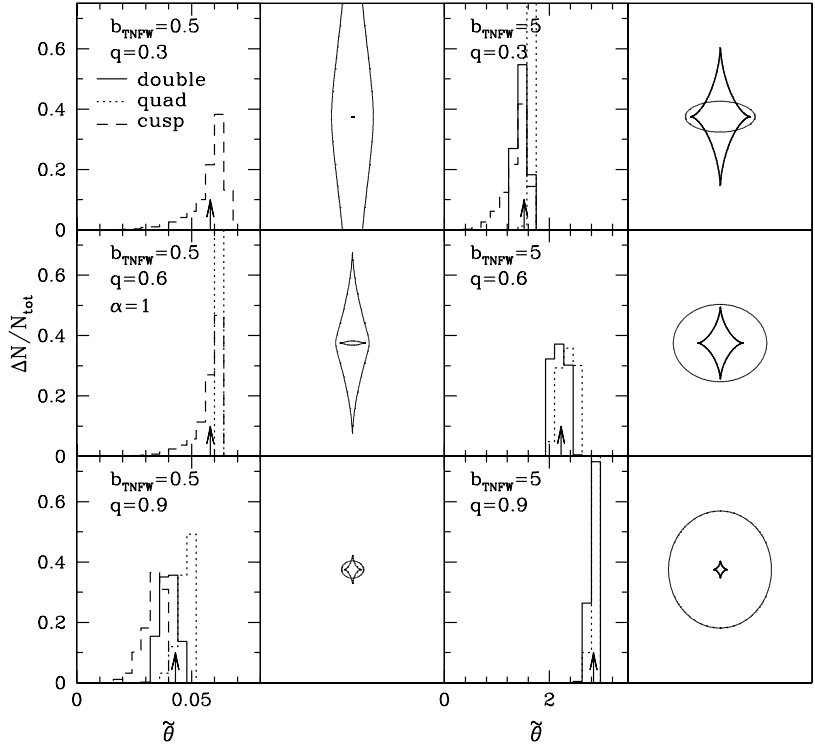}
\caption{Image separation distributions for sample lenses with
$\alpha=1$.  Arrows indicate the average separations.  The
corresponding caustics are shown for reference.  For each
$b_{\rm TNFW}$, the caustics are all plotted on the same scale.
\label{fig:dist_theta_a100}}
\end{figure*}

\begin{figure*}
\epsscale{0.7}
\plotone{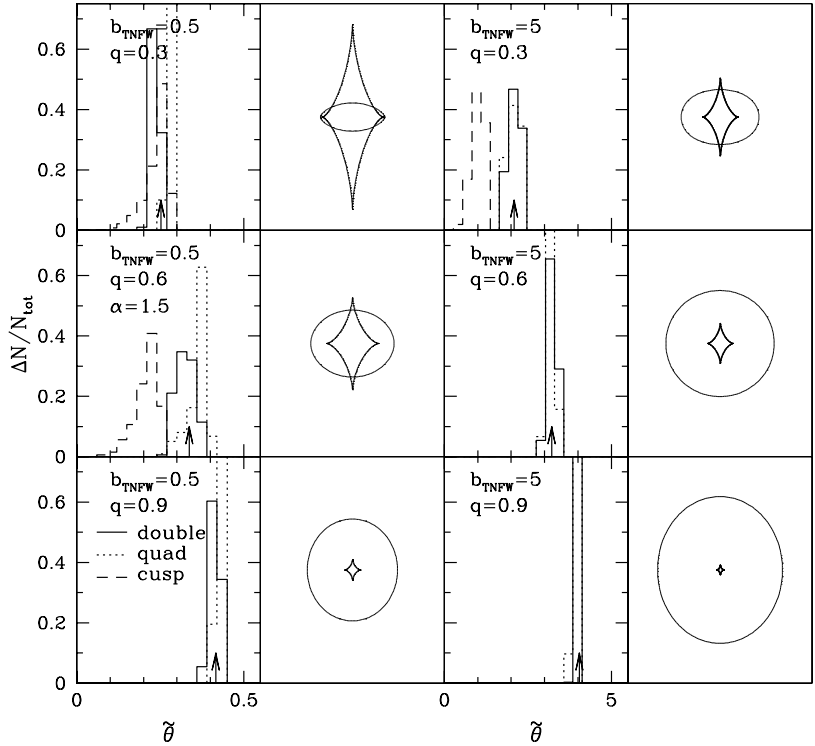}
\caption{Similar to Figure~\ref{fig:dist_theta_a100}, but for $\alpha=1.5$.
\label{fig:dist_theta_a150}}
\end{figure*}

We compute lensing cross sections using Monte Carlo methods.
Working in dimensionless coordinates $X \equiv x'/L_0$ and
$Y \equiv y'/L_0$, we pick random sources and use the
{\it gravlens} software by \citet{keeton01b} to solve the lens
equation.  Figure~\ref{fig:multi} shows examples of the three
different kinds of image configurations: double, quadruple, and
naked cusp lenses.\footnote{We use the terms ``double'' and
``quadruple'' because the third and fifth images are usually too
faint to be observed, although with the density profiles we use
here they are probably not as faint as for nearly-isothermal
lenses \citep[see][]{rusin02}.}  We count the number of sources
that produce lenses of different image multiplicities to
determine the dimensionless cross sections $\tilde{\sigma}_2$,
$\tilde{\sigma}_4$, and $\tilde{\sigma}_{\rm c}$ for doubles,
quadruples, and cusps, respectively.  For each set of images,
we define the dimensionless image separation $\tilde{\theta}$
to be the maximum separation between any pair of images; this
is a convenient definition that depends only on observable
quantities and is well defined for all image configurations
(no matter how many images there are).  We bin the sources by
the image separations they produce to derive image separation
distributions, as shown in Figures~\ref{fig:dist_theta_a100}
and \ref{fig:dist_theta_a150}.  For a given halo there is a
range of separations, but it tends to be fairly narrow
($\lesssim$20\%); the main exception is for cusp configurations,
which show a tail to small separations that corresponds to
sources near the cusp in the caustic.

If we not only count the sources but also weight them
appropriately, we can compute the magnification bias.
Specifically, if the sources have a simple power law luminosity
function $\phi_L(L)\propto L^{-\beta}$ then the ``biased cross
section'' can be written as
%%%%%%%%%%%%%%%%%%%%%%%%%%%
\begin{equation}
 B\tilde{\sigma} = \int dX dY \frac{\phi_L(L/\mu)/\mu}{\phi_L(L)}
 = \int dX dY \mu^{\beta-1},
\end{equation}
%%%%%%%%%%%%%%%%%%%%%%%%%%%
where the integral is over the multiply-imaged region of the
source plane.  We can compute the biased cross sections for
doubles, quadruples, and cusps similarly.  Each source is to be
weighted by $\mu^{\beta-1}$, where we take $\mu$ to be the
magnification of the second brightest image to reflect the method
of searching for large-separation lenses in observational data
such as the SDSS \citep[see][]{inada03,oguri04}.

%%%%%%%%%%%%%%%%%%%%%%%%%%%%%%%%%%%%%%%%%%%%%%%%%%%%%%%%%%%%%%%%%%%%%
%%%%%%%%%%%%%%%%%%%%%%%%%%%%%%%%%%%%%%%%%%%%%%%%%%%%%%%%%%%%%%%%%%%%%
%%%%%%%%%%%%%%%%%%%%%%%%%%%%%%%%%%%%%%%%%%%%%%%%%%%%%%%%%%%%%%%%%%%%%

An important qualitative result is already apparent from
Figure~\ref{fig:multi}.  CDM-type dark matter halos are very
sensitive to departures from spherical symmetry, in the sense
that even small projected ellipticities lead to large tangential
caustics and hence large quadruple cross sections.  When the
ellipticity is large, the tangential caustic is much larger than
the radial caustic and nearly all the images correspond to cusp
configurations.  This situation is notably different from what
happens in lensing by galaxies that have concentrated, roughly
isothermal mass distributions.  In that case, the ellipticity
must approach unity before cusp configurations become common
\citep[see][]{keeton97,rusin01}.  Such large ellipticities are
uncommon, and cusp configurations are correspondingly rare among
observed galaxy-scale lenses: among $\sim$80 known lenses there
is only one candidate \citep[APM~08279+5255;][]{lewis02}.  The
incidence of cusp configurations therefore appears to be a
significant distinction between normal and large-separation
lenses.

%%%%%%%%%%%%%%%%%%%%%%%%%%%%%%%%%%%%%%%%%%%%%%%%%%%%%%%%%%%
\subsection{Lensing probabilities}
%%%%%%%%%%%%%%%%%%%%%%%%%%%%%%%%%%%%%%%%%%%%%%%%%%%%%%%%%%%

The probability that a source at redshift $z_{\rm S}$ is lensed
into a system with image separation $\theta$ is computed by summing
the biased cross section over an appropriate population of lens
halos:
%%%%%%%%%%%%%%%%%%%%%%%%%%%
\begin{eqnarray}
 \frac{dP}{d\theta}(\theta,z_{\rm S})&=&
   \int dz_{\rm L}\frac{c\,dt}{dz_{\rm L}}(1+z_{\rm L})^3\int d(a/c)
   \int dc_e \nonumber\\
 && \times \int d(a/b) \int d\theta \int d\phi \biggl[ p(a/c)\,p(c_e)\nonumber\\
 && \hspace*{7mm}p(a/b|a/c)\,p(\theta)\,p(\phi)\,B\sigma
   \,\frac{dn}{dM} \biggr]_{M(\theta)}.
   \label{eq:prob}
\end{eqnarray}
%%%%%%%%%%%%%%%%%%%%%%%%%%%
The first integral is over the volume between the observer and the
source.  The next three integrals are over the structural parameters
of the lens halos, while the last two integrals cover the different
orientations.  The mass function of dark matter halos is $dn/dM$,
and we use the model from equation (B3) of \citet{jenkins01}.  Finally,
$M(\theta)$ is the mass of a halo that produces image separation
$\theta$ (for given redshift and other parameters), which is given
by the solution of
%%%%%%%%%%%%%%%%%%%%%%%%%%%
\begin{equation}
 \theta = R_0\,q_x\,\tilde{\theta}(b_{\rm TNFW},q).
\end{equation}
%%%%%%%%%%%%%%%%%%%%%%%%%%%
The square brackets in equation (\ref{eq:prob}) indicate that the
integrand is to be evaluated only for parameter sets that produce
the desired image separation.  Equation (\ref{eq:prob}) gives the
total lensing probability, but we can simply replace the total
biased cross section $B \sigma$ with $B_2 \sigma_2$, $B_4 \sigma_4$,
or $B_c \sigma_c$ to compute the probability for doubles, quadruples,
or cusps.

We use the model given by JS02 for the probability distribution
functions (PDFs) that appear in equation (\ref{eq:prob}).  The PDFs for
the axis ratios are:
%%%%%%%%%%%%%%%%%%%%%%%%%%%
\begin{eqnarray}
 p(a/c)=\frac{1}{\sqrt{2\pi}\times 0.113}&&\nonumber\\
&&\hspace*{-30mm}\times\exp\left[-\frac{\left\{(a/c)(M_{\rm vir}/M_*)^{0.07[\Omega(z)]^{0.7}}-0.54\right\}^2}{2(0.113)^2}\right]\nonumber\\
&&\hspace*{-30mm}\times\left(\frac{M_{\rm vir}}{M_*}\right)^{0.07[\Omega(z)]^{0.7}},
\label{p_a}
\end{eqnarray}
%%%%%%%%%%%%%%%%%%%%%%%%%%%
%%%%%%%%%%%%%%%%%%%%%%%%%%%
\begin{eqnarray}
 p(a/b|a/c)=\frac{3}{2(1-\max(a/c,0.5))}&&\nonumber\\
&&\hspace*{-40mm}\times\left[1-\left(\frac{2a/b-1-\max(a/c,0.5)}{1-\max(a/c,0.5)}\right)^2\right],
\label{p_ab}
\end{eqnarray}
%%%%%%%%%%%%%%%%%%%%%%%%%%%
where $M_*$ is the characteristic nonlinear mass such that the
RMS top-hat-smoothed overdensity at that mass scale is $1.68$.
Note that $M_*$ and $\Omega(z)$ depend on redshift, so $p(a/c)$ varies
with redshift such that halos tend to be more triaxial at higher
redshifts (see Fig.~8 of JS02).  Also note that $p(a/b|a/c)=0$
for $a/b < \max(a/c,0.5)$.  The PDF for the concentration
parameter is
%%%%%%%%%%%%%%%%%%%%%%%%%%%
\begin{equation}
 p(c_e)=\frac{1}{\sqrt{2\pi} \times 0.3}
\exp\left[-\frac{(\ln c_e-\ln \bar{c}_e)^2}{2(0.3)^2}\right]\frac{1}{c_e}.
\label{p_ce}
\end{equation}
%%%%%%%%%%%%%%%%%%%%%%%%%%%
For the median concentration parameter $\bar{c}_e$, we adopt a
fitting formula given by JS02 (see also OLS03):
%%%%%%%%%%%%%%%%%%%%%%%%%%%
\begin{eqnarray}
 \bar{c}_e=1.35\exp\left[-\left\{\frac{0.3}{(a/c)(M_{\rm vir}/M_*)^{0.07[\Omega(z)]^{0.7}}}\right\}^2\right]&&\nonumber\\
&&\hspace*{-50mm}\times A_e\sqrt{\frac{\Delta_{\rm vir}(z_c)}{\Delta_{\rm vir}(z)}}\left(\frac{1+z_c}{1+z}\right)^{3/2},
\label{ce}
\end{eqnarray}
%%%%%%%%%%%%%%%%%%%%%%%%%%%
with $z_c$ being the collapse redshift of the halo of mass
$M_{\rm vir}$ and $A_e=1.1$ in the Lambda-dominated CDM model (JS02).
For spherical halos, large-separation lens statistics are highly
sensitive to the concentration distribution \citep[both the median
and the scatter; see][]{kuhlen04}.  However, we checked that in
the triaxial model the sensitivity is much reduced because there
is such a broad distribution of axis ratios; hence we do not present
results with a different median or scatter in the concentration
distribution.  Finally, the PDFs for the orientation angles are
%%%%%%%%%%%%%%%%%%%%%%%%%
\begin{eqnarray}
 p(\theta) &=& \frac{\sin\theta}{2}, \label{p_theta} \\
 p(\phi)   &=& \frac{1}{2\pi},
\end{eqnarray}
%%%%%%%%%%%%%%%%%%%%%%%%%
corresponding to random 3-d orientations.

%%%%%%%%%%%%%%%%%%%%%%%%%%%%%%%%%%%%%%%%%%%%%%%%%%%%%%%%%%%
%%%%%%%%%%%%%%%%%%%%%%%%%%%%%%%%%%%%%%%%%%%%%%%%%%%%%%%%%%%
\section{Lensing Probabilities and Image Multiplicities in the
Triaxial Halo Model}\label{sec:prob}
%%%%%%%%%%%%%%%%%%%%%%%%%%%%%%%%%%%%%%%%%%%%%%%%%%%%%%%%%%%
%%%%%%%%%%%%%%%%%%%%%%%%%%%%%%%%%%%%%%%%%%%%%%%%%%%%%%%%%%%

%%%%%%%%%%%%%%%%%%%%%%%%%%%%%%%%%%%%%%%%%%%%%%%%%%%%%%%%%%%
\subsection{Dependence of the triaxiality}
%%%%%%%%%%%%%%%%%%%%%%%%%%%%%%%%%%%%%%%%%%%%%%%%%%%%%%%%%%%

%%%%%%%%%%%%%%%%%%%%%%%%%%%%%%%%%%%%%%%%%%%%%%%%%%%%%%%%%%%%%%%%%%%%%
\begin{figure}
\epsscale{1.15}
\plotone{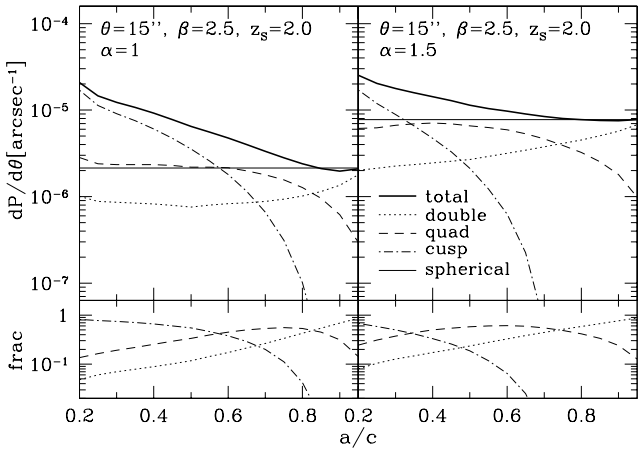}
\caption{Lensing probabilities and image multiplicities as a function
of $a/c$.  The left panels are for inner slope $\alpha=1$, and the
right panels for $\alpha=1.5$.  We adopt $\theta=15''$, $\beta=2.5$,
and $z_{\rm S}=2.0$.  We show the total lensing probability (thick
solid line), as well as the probabilities for double (dotted),
quadruple (dashed), and cusp (dash-dotted) lenses.  For comparison,
the lensing probability of spherical halos is shown by the thin solid
line.
\label{fig:afix}}
\end{figure}
%%%%%%%%%%%%%%%%%%%%%%%%%%%%%%%%%%%%%%%%%%%%%%%%%%%%%%%%%%%%%%%%%%%%%

We begin by examining how the lensing probabilities and image
multiplicities vary when we change the degree of triaxiality.
We remove the integral over $a/c$ in equation (\ref{eq:prob}) to
compute the lensing probabilities at fixed triaxiality (we still
integrate over the intermediate axis ratio $a/b$ and over random
orientations).  We can then plot the probabilities as a function
of $a/c$, as shown in Figure~\ref{fig:afix}.  In this example, we
place the source at redshift $z_{\rm S}=2.0$, and we use a source
luminosity function with slope $\beta=2.5$.  We compute the
probabilities for an image separation of $\theta=15''$, similar
to that of the one known large-separation lensed quasar
SDSS J1004+4112 \citep{inada03}.

For $a/c \to 1$ we recover the spherical case.  As $a/c$ decreases
(the triaxiality increases), at first the total lensing probability
stays roughly constant but the fraction of quadruples rises; this is
similar to the effects of ellipticity on isothermal lenses
\citep[see][]{keeton97,rusin01}.  Then the probability for naked
cusp image configurations begins to rise dramatically, and they
come to dominate the total probability.  Interestingly, the sum of
the probabilities for quadruple and double lenses is roughly equal
to the probability for spherical halos for most values of $a/c$,
at least for this example with image separation $\theta=15''$ (also
see Figures~\ref{fig:sep}--\ref{fig:zs} below).  This suggests that
the enhancement in the total lensing probability is mainly driven
by naked cusp configurations.  The $\alpha=1$ and 1.5 cases both
have these qualitative features, and they differ only in the
quantitative details.  Since the typical triaxiality in CDM
simulations is $a/c \sim 0.5$ (JS02), it appears that triaxiality
can have a significant effect on the statistics of large-separation
lenses.

%%%%%%%%%%%%%%%%%%%%%%%%%%%%%%%%%%%%%%%%%%%%%%%%%%%%%%%%%%%
\subsection{Full results}
%%%%%%%%%%%%%%%%%%%%%%%%%%%%%%%%%%%%%%%%%%%%%%%%%%%%%%%%%%%

%%%%%%%%%%%%%%%%%%%%%%%%%%%%%%%%%%%%%%%%%%%%%%%%%%%%%%%%%%%%%%%%%%%%%
\begin{figure}
\epsscale{1.15}
\plotone{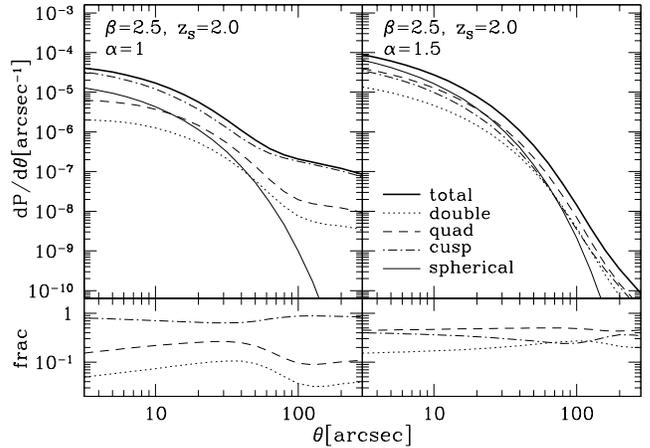}
\caption{Lensing probabilities and image multiplicities with triaxial
dark halos as a function of image separation $\theta$.  The source is
placed at $z_{\rm S}=2.0$, and slope of the source luminosity function
is fixed to $\beta=2.5$.
\label{fig:sep}}
\end{figure}
%%%%%%%%%%%%%%%%%%%%%%%%%%%%%%%%%%%%%%%%%%%%%%%%%%%%%%%%%%%%%%%%%%%%%

To compute the full impact of triaxiality on lens statistics, we
must integrate over an appropriate triaxiality distribution (as in
equation \ref{eq:prob}).  Figure~\ref{fig:sep} shows the resulting
lensing probabilities and image multiplicities as a function of
the image separation $\theta$.  Again, we place the source quasar
at $z_{\rm S}=2.0$, and fix the slope of the source luminosity
function to $\beta=2.5$.  The first important result is that the
triaxial model predicts larger lensing probabilities than the
spherical model for all image separations.  The enhancement is a
factor of $\sim$4 for $\alpha=1$, and a factor of $\sim$2 for
$\alpha=1.5$, if the image separation is not so large
($\theta\lesssim 30''$).  At larger separations it seems that the
effect of triaxiality is even more significant, especially for
$\alpha=1$; we will discuss this issue in \S\ref{sec:largesep}.

There are several interesting results in the image multiplicities.
The $\alpha=1$ and $1.5$ cases have very different multiplicities:
with $\alpha=1$ the lensing probability is dominated by cusp
configurations; while with $\alpha=1.5$ quadruple lenses are
somewhat more common than cusps.  Neither result is very sensitive
to the image separation.  In both cases double lenses are fairly
uncommon, which is very different from the situation with normal
arcsecond-scale lenses produced by nearly-isothermal galaxies.
This result is consistent with previous theoretical conclusions
that image multiplicities depend on the central concentration of
the lens mass distribution, such that less concentrated profiles
tend to produce more quadruple and cusp lenses
(\citealt{kassiola93}; \citealt*{kormann94}; \citealt{rusin01};
\citealt{evans02}; \citealt{dalal04}).
The important point for observations is that if dark halos have
inner profiles with $\alpha \lesssim 1.5$, then many or even most
large-separation lenses should be quadruples or cusps rather than
doubles.  (We will consider the implications for SDSS J1004+4112
in \S\ref{sec:sdss}.)  Another point is that the image
multiplicities are sensitive to the inner density profile, so they
offer a new method for probing dark matter density profiles that
is qualitatively different from methods discussed before.

%%%%%%%%%%%%%%%%%%%%%%%%%%%%%%%%%%%%%%%%%%%%%%%%%%%%%%%%%%%%%%%%%%%%%
\begin{figure}
\epsscale{1.15}
\plotone{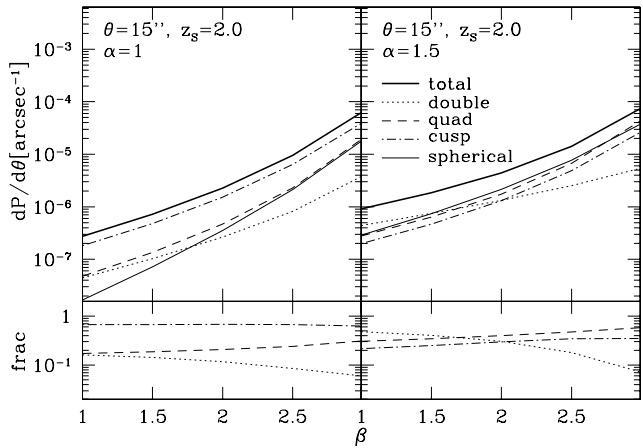}
\caption{Lensing probabilities and image multiplicities with triaxial
dark halos as a function of the slope of the source luminosity function
$\beta$.  We consider an image separation of $\theta=15''$, and we
place the source at $z_{\rm S}=2.0$.
 \label{fig:bias}}
\end{figure}
%%%%%%%%%%%%%%%%%%%%%%%%%%%%%%%%%%%%%%%%%%%%%%%%%%%%%%%%%%%%%%%%%%%%%

Magnification bias is important in lens statistics, particularly
in image multiplicities, because it gives more weight to quadruple
and cusp configurations (which tend to have large magnifications)
than to doubles.  We should therefore understand what happens when
we modify the magnification bias by varying the slope $\beta$ of
the source luminosity function.  The results are shown in
Figure~\ref{fig:bias} (for image separation $\theta=15''$ and
source redshift $z_{\rm S}=2.0$).  The lensing probabilities
increase as $\beta$ increases, because as the source luminosity
function becomes steeper magnification bias becomes
stronger.\footnote{Note that the magnification bias diverges if
the luminosity function is a pure power law with $\beta \ge 3$,
so we are restricted to shallower cases.}  Interestingly, the
increase in the total probability due to triaxiality weakens as
$\beta$ increases, although the effect is not strong.  As for
the image multiplicities, $\alpha=1$ halos are always dominated
by cusp lenses, although for sufficiently steep luminosity
functions quadruples become fairly common.  With $\alpha=1.5$
halos, when magnification bias is weak ($\beta \sim 1$) doubles
are the most probable, but as magnification bias strengthens
($\beta$ increases) quadruples receive more weight and become
the most likely.  In practice, the effective values of $\beta$
are larger than $\sim$1.5 for both optical \citep[e.g.,][]{boyle00}
and radio surveys \citep[e.g.,][]{rusin01}, so we expect cusps
to dominate for $\alpha=1$ and quadruples to be the most common
for $\alpha=1.5$.

%%%%%%%%%%%%%%%%%%%%%%%%%%%%%%%%%%%%%%%%%%%%%%%%%%%%%%%%%%%%%%%%%%%%%
\begin{figure}
\epsscale{1.15}
\plotone{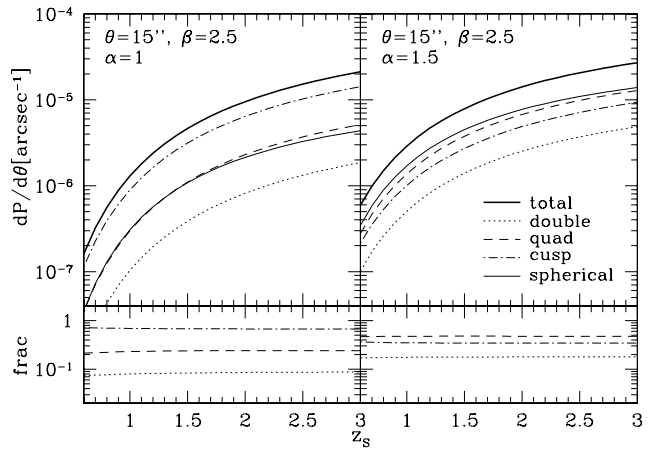}
\caption{Lensing probabilities and image multiplicities as a function
of source redshift $z_{\rm S}$.  We fix the image separation to
$\theta=15''$ and the of the source luminosity function to $\beta=2.5$.
\label{fig:zs}}
\end{figure}
%%%%%%%%%%%%%%%%%%%%%%%%%%%%%%%%%%%%%%%%%%%%%%%%%%%%%%%%%%%%%%%%%%%%%

Finally, we consider whether the results depend on the source
redshift, as shown in Figure~\ref{fig:zs}.  The lensing probabilities
rise with $z_{\rm S}$, because there is more volume and hence
more deflectors between the observer and the source.  In addition,
geometric effects mean that a given halo can produce a larger
image separation when the source is more distant, so the halo
mass required to produce a given image separation goes down and
the abundance of relevant deflectors goes up.  However, the
probability increase affects the different image configurations
in basically the same way, so the image multiplicities are quite
insensitive to the source redshift.  Therefore, we conclude that
details of the source redshift distribution are not so important
for image multiplicities, at least when the source luminosity
function has a power law shape.

%%%%%%%%%%%%%%%%%%%%%%%%%%%%%%%%%%%%%%%%%%%%%%%%%%%%%%%%%%%
\subsection{Statistics at larger image separations}\label{sec:largesep}
%%%%%%%%%%%%%%%%%%%%%%%%%%%%%%%%%%%%%%%%%%%%%%%%%%%%%%%%%%%

In Figure~\ref{fig:sep}, at very large image separations
($\theta\gtrsim100''$) the lensing probabilities in the triaxial
halo model are orders of magnitude larger than those in the
spherical halo model. In addition, at these very large separations
the $\alpha=1$ case produces higher probabilities than the more
concentrated $\alpha=1.5$ case.  Both features are puzzling and
invite careful consideration.

%%%%%%%%%%%%%%%%%%%%%%%%%%%%%%%%%%%%%%%%%%%%%%%%%%%%%%%%%%%%%%%%%%%%%
\begin{figure}
\epsscale{0.85}
\plotone{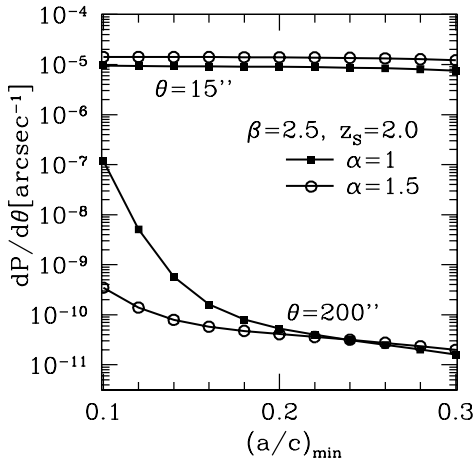}
\caption{Dependence of the lensing probability on the cutoff in $a/c$.
Both $\theta=15''$ and $200''$ are shown.  Filled squares and open
circles denote $\alpha=1$ and $1.5$, respectively.  The slope of the
luminosity function is fixed to $\beta=2.5$.
\label{fig:amin}}
\end{figure}
%%%%%%%%%%%%%%%%%%%%%%%%%%%%%%%%%%%%%%%%%%%%%%%%%%%%%%%%%%%%%%%%%%%%%

Figure~\ref{fig:amin} shows the dependence of total lensing
probability on the lower limit of the integral over $a/c$.  In
the previous calculations, we assumed $(a/c)_{\rm min}=0.1$.
This figure shows that for $\theta=15''$ the results are quite
insensitive to $(a/c)_{\rm min}$, suggesting that the
contribution from extremely triaxial halos is negligible.
For $\theta=200''$, however, the lensing probability rapidly
decreases as $(a/c)_{\rm min}$ increases. In other words, the
lensing probability at very large image separations seems to be
dominated by very small $a/c$, or very large triaxialities.

Results that are dominated by such extreme halos are probably not
very reliable. They depend sensitively on both the assumed PDF for
the axis ratio $a/c$ (eq.~[\ref{p_a}]) and the correlation between
$a/c$ and the concentration $c_e$ (eq.~[\ref{ce}]) at very small
axis ratios. The fitting forms presented by JS02 were intended to
reproduce the PDF and correlation at $a/c\gtrsim 0.3$, and it is
unclear whether they are still accurate at $a/c \sim 0.1$. In
addition, even if we know accurate fitting forms, such a situation
implies that sample variance (i.e., the effect of the finite
number of lensing clusters) may be quite large.

Another ambiguity is the projection effect. In this paper, we
assumed that the density profile (eq.~[\ref{gnfw}]) extends
beyond the virial radius, and in projecting along the line of
sight we integrated the profile to infinity. Although it is not
clear whether we should cut off the profile at the virial radius
or not \citep[e.g.,][]{takada03}, the effect of the extended
profile on the gravitational lensing is not so large for normal
dark halos. However, when $a/c$ is small enough, equation (\ref{ce})
indicates that the concentration parameter $c_e$ becomes smaller
than unity, so the effect of the extended profile outside the
virial radius is quite significant. The puzzling feature that the
$\alpha=1$ case produces higher probabilities than the
$\alpha=1.5$ case at very large separations can be ascribed to the
projection effect, because the effect is more significant for
shallower density profiles.

Thus, lenses with extremely large image separations are associated
with the most extreme dark matter halos, and it may be difficult
to make reliable predictions about them.  We emphasize, though,
that these issues do not apply to lenses with separations $\theta
\lesssim 30''$, and on these scales we believe our results to be
robust.

%%%%%%%%%%%%%%%%%%%%%%%%%%%%%%%%%%%%%%%%%%%%%%%%%%%%%%%%%%%
%%%%%%%%%%%%%%%%%%%%%%%%%%%%%%%%%%%%%%%%%%%%%%%%%%%%%%%%%%%
\section{Predictions for the SDSS}\label{sec:sdss}
%%%%%%%%%%%%%%%%%%%%%%%%%%%%%%%%%%%%%%%%%%%%%%%%%%%%%%%%%%%
%%%%%%%%%%%%%%%%%%%%%%%%%%%%%%%%%%%%%%%%%%%%%%%%%%%%%%%%%%%

In the previous section, we assumed a simple power law source
luminosity function to understand the general effects of
triaxiality. Here we consider a specific luminosity function
appropriate for quasars, and further customize our predictions to
quasars in the Sloan Digital Sky Survey.

We adopt the double power law $B$-band luminosity function for
quasars proposed by \citet*{boyle88},
%%%%%%%%%%%%%%%%%%%%%%%%%%%%%%%%%%%%%%%%%%%%%%%%%%%%%%%%%%%%%%%%%%%%%%%%
\begin{equation}
 \phi_L(z_{\rm S},L)dL=\frac{\phi_*}{[L/L_*(z_{\rm S})]^{\beta_l}+[L/L_*(z_{\rm S})]^{\beta_h}}\frac{dL}{L_*(z_{\rm S})},
\end{equation}
%%%%%%%%%%%%%%%%%%%%%%%%%%%%%%%%%%%%%%%%%%%%%%%%%%%%%%%%%%%%%%%%%%%%%%%%
and use $\beta_h=3.43$ and $\beta_l=1.64$ \citep[see][]{boyle00}.
We let the break luminosity evolve following a model that
reproduces the low-redshift luminosity function as well as the
space density of high-redshift quasars
\citep[see][]{wyithe02,oguri04}. In practice we actually use the
cumulative luminosity function
%%%%%%%%%%%%%%%%%%%%%%%%%%%%%%%%%%%%%%%%%%%%%%%%%%%%%%%%%%%%%%%%%%%%%%%%
\begin{equation}
\Phi_L(z_{\rm S},L)=\int^\infty_L\phi_L(z_{\rm S},L)dL,
\end{equation}
%%%%%%%%%%%%%%%%%%%%%%%%%%%%%%%%%%%%%%%%%%%%%%%%%%%%%%%%%%%%%%%%%%%%%%%%
to calculate the biased cross section (see \S\ref{sec:cross})
%%%%%%%%%%%%%%%%%%%%%%%%%%%
\begin{equation}
 B\tilde{\sigma}=\int dXdY \frac{\Phi_L(L/\mu)}{\Phi_L(L)}.
\end{equation}
%%%%%%%%%%%%%%%%%%%%%%%%%%%
We approximate that the SDSS quasar sample is a sample with a flux
limit of $i^*=19.7$.\footnote{The SDSS quasar target selection is
aimed to choose quasars with $i^*\lesssim 19.1$ \citep{richards02}.
However, we assume a flux limit of $i^*=19.7$ because there are
quasars with $i^*>19.1$ in the SDSS quasar sample which are, for
example, first targeted as different objects but revealed to be
quasars.}  One needs a cross-filter K-correction to convert
observed $i^*$ magnitudes to absolute $B$-band luminosity.  We
adopt the following form
%%%%%%%%%%%%%%%%%%%%%%%%%%%%%%%%%%%%%%%%%%%%%%%%%%%%%%%%%%%%%%%%%%%%%%%%
\begin{eqnarray}
 K_{Bi}(z)&=&-2.5(1-\alpha_{\rm s})\log(1+z)\nonumber\\
&&-2.5\alpha_{\rm s}\log\left(\frac{7500}{4400}\right)-0.12,
\label{kcor_approx}
\end{eqnarray}
%%%%%%%%%%%%%%%%%%%%%%%%%%%%%%%%%%%%%%%%%%%%%%%%%%%%%%%%%%%%%%%%%%%%%%%%
with $\alpha_{\rm s}=0.5$ \citep{oguri04}. Finally, we approximate
the redshift distribution (see Fig.~1 of \citealt{oguri04}) with
the following Gaussian distribution:
%%%%%%%%%%%%%%%%%%%%%%%%%%%
\begin{eqnarray}
p(z_{\rm S})dz_{\rm S}=\frac{1}{1.21}\exp\left\{-\frac{(z-1.45)^2}{2(0.55)^2}\right\}dz_{\rm S}&&\nonumber\\ 
&&\hspace*{-20mm}\mbox{($0.6<z_{\rm S}<2.3$)}.
\label{sdss_zdist}
\end{eqnarray}
%%%%%%%%%%%%%%%%%%%%%%%%%%%
We have confirmed that the results using these approximations
agree well with those obtained by fully taking account of the
observed redshift and magnitude distributions.

%%%%%%%%%%%%%%%%%%%%%%%%%%%%%%%%%%%%%%%%%%%%%%%%%%%%%%%%%%%%%%%%%%%%%
\begin{figure}
\epsscale{1.15}
\plotone{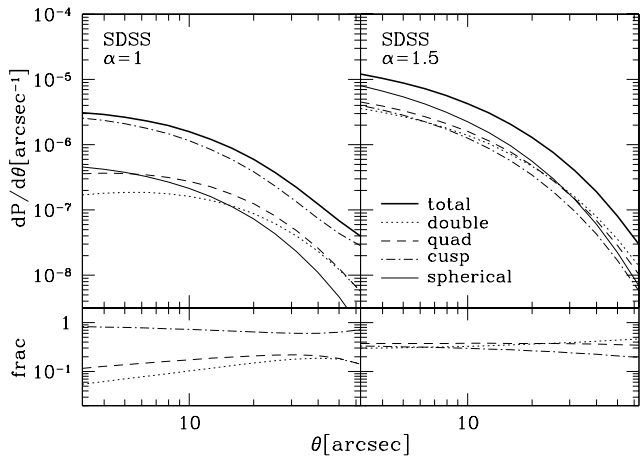}
\caption{Lensing probabilities and image multiplicities for SDSS
quasars at redshifts $0.6<z_{\rm S}<2.3$.
\label{fig:sdss_sep}}
\end{figure}
%%%%%%%%%%%%%%%%%%%%%%%%%%%%%%%%%%%%%%%%%%%%%%%%%%%%%%%%%%%%%%%%%%%%%

Figure~\ref{fig:sdss_sep} shows the lensing probabilities and
image multiplicities as a function of image separation $\theta$
for the SDSS quasar sample.  Again large-separation lenses are
dominated by cusp configurations for $\alpha=1$, but all three
configurations are almost equally likely for $\alpha=1.5$.
Therefore, we confirm that image multiplicities in SDSS
large-separation lenses will offer interesting information on the
density profile of dark halos.  We can now consider whether it is
statistically natural that the first large-separation lens in the
SDSS is a quadruple lens.  We find that for an image separation
$\theta=15''$ the fractions of quadruple lenses are $\sim$0.2 and
$\sim$0.4 for $\alpha=1$ and $1.5$, respectively. Thus $\alpha=1.5$
could explain the discovery of the quadruple lens somewhat better,
but $\alpha=1$ is also not unnatural.

Finally, we can use the discovery of SDSS J1004+4112, together
with the lack of large-separation lenses in the Cosmic Lens
All-Sky Survey \citep[CLASS;][]{phillips01}, to constrain the
cosmological parameter $\sigma_8$ describing the normalization
of the density fluctuation power spectrum.  Using spherical
models, \citet{oguri04} found that the discovery of SDSS J1004+4112
required rather large values of either $\alpha$ or $\sigma_8$,
but given the importance of triaxiality we should revisit this
question. For the SDSS, we compute the expected number of
large-separation lenses with $7''<\theta<60''$ among the 29811
SDSS quasars.  For CLASS, we adopt a power law source luminosity
function with $\beta=2.1$ \citep[see][]{rusin01}, fix the source
redshift to $z_{\rm S}=1.3$ \citep[see][]{marlow00}, and calculate
the expected number of lenses with $6''<\theta<15''$ among 9284
flat-spectrum radio sources \citep{phillips01}.  We then compute
the likelihood
%%%%%%%%%%%%%%%%%%%%%%%%%%%%%%%%%%%%%%%%%%%%%%%%%%%%%%%%%%%%%%%%%%%%%%%%
\begin{equation}
\mathcal{L}\propto \left(1-e^{-N_{\rm SDSS}}\right) e^{-N_{\rm CLASS}},
\end{equation}
%%%%%%%%%%%%%%%%%%%%%%%%%%%%%%%%%%%%%%%%%%%%%%%%%%%%%%%%%%%%%%%%%%%%%%%%
which represents the Poisson probability of observing no
large-separation lenses in CLASS when $N_{\rm CLASS}$ are expected,
and at least one large-separation lens in SDSS when $N_{\rm SDSS}$
are expected.  (There may be other large-separation lenses in the
SDSS sample that have not yet been identified.)  We consider two
possibilities for the expected number of lenses in the SDSS:
(1) the total number of lenses $N_{\rm tot}$ is used as
$N_{\rm SDSS}$; (2) only the number of quadruple lenses
$N_{\rm quad}$ is because the discovered lens is quadruple.

%%%%%%%%%%%%%%%%%%%%%%%%%%%%%%%%%%%%%%%%%%%%%%%%%%%%%%%%%%%%%%%%%%%%%
\begin{figure}
\epsscale{0.85}
\plotone{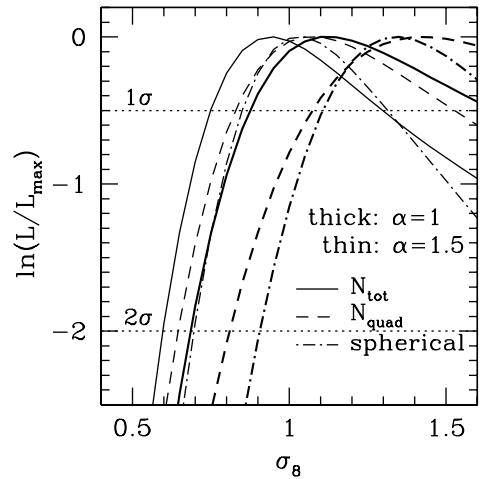}
\caption{Maximum likelihood estimates for $\sigma_8$ from combining
the discovery of SDSS J1004+4112 in SDSS with the lack of
large-separation lenses in CLASS.  In making predictions for SDSS,
we consider two cases: the appropriate prediction could be the
total number of lenses (solid lines); or since SDSS J1004+4112 is
a quad the appropriate quantity could be the number of quadruples
(dashed lines).  The likelihoods for $\alpha=1$ and $1.5$ are shown
by thick and thin lines, respectively.  Results for spherical halos
are also shown by dash-dotted lines for reference.
\label{fig:s8}}
\end{figure}
%%%%%%%%%%%%%%%%%%%%%%%%%%%%%%%%%%%%%%%%%%%%%%%%%%%%%%%%%%%%%%%%%%%%%

Figure~\ref{fig:s8} shows the resulting maximum likelihood
constraints on $\sigma_8$.  We find that $\sigma_8 \sim 1$
explains the data well, although the details depend on the value
of $\alpha$ and the choice of $N_{\rm SDSS}$. Only the case with
$\alpha=1$ and $N_{\rm SDSS}=N_{\rm quad}$ prefers relatively
large $\sigma_8$, but $\sigma_8=1$ is still allowed at the
2$\sigma$ level. At present the data do not allow particularly
strong constraints on $\sigma_8$.  Nevertheless, we can conclude
that the status large-separation lenses is quite consistent with
the predictions of CDM given $\sigma_8 \sim 1$.

For comparison, Figure~\ref{fig:s8} also shows results for spherical
halos.  It turns out that the spherical model overestimates the
value of $\sigma_8$ by $\sim$0.1 for $\alpha=1$ and $\sim$0.2 for
$\alpha=1.5$, compared with cases where we take $N_{\rm SDSS} = N_{\rm tot}$.
Triaxiality is therefore an important systematic effect in these
cases.  Interestingly, the best-fit values of $\sigma_8$ from
spherical models are quite similar to those from triaxial models
with $N_{\rm SDSS} = N_{\rm quad}$.  In both cases, the
likelihood function for the spherical model is narrower than for
the triaxial model, indicating that the spherical model would
underestimate the statistical uncertainties in $\sigma_8$.

%%%%%%%%%%%%%%%%%%%%%%%%%%%%%%%%%%%%%%%%%%%%%%%%%%%%%%%%%%%
%%%%%%%%%%%%%%%%%%%%%%%%%%%%%%%%%%%%%%%%%%%%%%%%%%%%%%%%%%%
\section{Summary and Discussion}\label{sec:sum}
%%%%%%%%%%%%%%%%%%%%%%%%%%%%%%%%%%%%%%%%%%%%%%%%%%%%%%%%%%%
%%%%%%%%%%%%%%%%%%%%%%%%%%%%%%%%%%%%%%%%%%%%%%%%%%%%%%%%%%%

The dark matter halos predicted by the CDM model are triaxial
rather than spherical, which has a significant effect on the
statistics of large-separation gravitational lenses.  Triaxiality
systematically enhances the lensing probability by a factor of
$\sim$4 if dark halos have an inner density profile with
$\alpha=1$, or a factor of $\sim$2 if $\alpha=1.5$.  The effects
may be even more dramatic at very large image separations
($\theta \gtrsim 100''$), although such lenses are very sensitive
to the most triaxial halos and so the predictions are not as
reliable.  Thus, triaxiality must be added to the list of
important systematic effects that need to be included in
calculations of large-separation lens statistics.  (Some of the
other effects are the inner density profile and the shape of the
distribution of concentration parameters, as found by previous
studies of large-separation lens statistics using spherical
halos.)

Triaxial modeling allows us to predict the image multiplicities
for large-separation lenses.  We found that the multiplicities
depend strongly on the density profile: for $\alpha=1$, lenses
are dominated by naked cusp image configurations; while for
$\alpha=1.5$, quadruple configurations are the most probable.
Double lenses, which are dominant among normal arcsecond-scale
lenses, are subdominant in both cases. Note that cusp lenses can be 
distinguished from doubles by the presence of a third image
comparable in brightness to the other two, and by the configuration
of image positions.  The differences can be ascribed to the
different mass density profiles, and they indicate that the
multiplicities of large-separation lenses will provide a
qualitatively new probe of the central density profiles of
massive dark matter halos (and hence a new test of CDM).

We have computed lensing probabilities and image multiplicities
for the SDSS quasar sample.  We predict that for both $\alpha=1$
and $1.5$ most of the large-separation lenses should be quadruples
or cusps.  The fractions of quadruple lenses at separations of
$\theta=15''$ are $\sim$0.2 and $\sim$0.4 for $\alpha=1$ and $1.5$,
respectively.  Thus it is not surprising that the first
large-separation lens discovered is a quadruple.  In addition, we
computed the expected number of large-separation lenses in both
SDSS and CLASS, and found that the data are consistent with the
CDM model with $\sigma_8 \sim 1$, in agreement with other
measurements \citep[e.g.,][and references therein]{spergel03}.
Thus, the discovery of SDSS J1004+4112 can be interpreted as
additional support for CDM on non-linear cluster scales.

\lastpagefootnotes
The prediction that triaxial halos produce significant fractions
of quadruple and cusp lenses should be kept in mind when
considering samples of candidate large-separation lenses.  For
example, \citet{miller04} found six large-separation double lens
candidates in the 2dF Quasar Redshift Survey, but no quadruple or
cusp lens candidates.  Even accounting for small number statistics,
our results suggest that such a high fraction of doubles would be
inconsistent with CDM at more than 3$\sigma$,\footnote{If the
predicted double fraction is $f_2$, then the Poisson probability
of having $N$ doubles and no quadruples or cusps is
$\mathcal{L}(f_2) \propto (f_2)^N$.} and that it would be
surprising if many of the six candidates are genuine lens systems.

We note that the triaxial dark halo model we adopted in this paper
can be improved in several ways. First, we assumed that the axis
ratios of the triaxial ellipsoids are constant with radius. However,
JS02 showed that the axis ratios decrease slightly toward the halo
centers: $a/c$ decreases by $\sim\!0.2$ as the mean radius decreases
from $\sim\!0.6 r_{\rm vir}$ to $\sim\!0.06 r_{\rm vir}$.  Since
strong lensing is most sensitive to the inner parts of dark halos,
it is possible that we have actually underestimated the effects of
triaxiality on the statistics of large-separation lenses.  On the
other hand, including baryons (which were neglected in the
simulations of JS02) would tend to make dark halos rounder by
$\Delta(a/c) \sim 0.1$--$0.2$ because of the isotropic gas pressure. 

While our theoretical model is much more realistic than the
simple spherical model, we have still made several simplifying
assumptions.  One is that we have neglected substructure in dark
halos.  The galaxies in massive cluster halos do not have a large
effect on the statistics of lensed arcs \citep{meneghetti00}, but
it is not obvious whether or not they would affect large-separation
lenses.  Substructure can affect the image multiplicities for
isothermal lenses \citep{cohn04}, so it should be considered for
CDM halos as well.  Another effect we have neglected is the presence
of a massive central galaxy in a cluster.  \citet*{meneghetti03b}
claim that central galaxies do not have a large effect on arc
statistics.  However, because our results depend on the inner
slope of the density profile, and a central galaxy effectively
increases the concentration, this effect should be considered.
A third phenomenon we have neglected is cluster merger events.
Indeed, mergers can change the shapes of critical curves and caustics
substantially, and thus have a great impact on lensing cross sections
\citep{torri04}.  To estimate the effect on large-separation lens
statistics, we would need a realistic model of the cluster merger
event rate and the physical conditions of merger events.  Addressing
these various issues to improve the accuracy of the theoretical
predictions is beyond the scope of this paper, but is certainly of
interest for future work.

%%%%%%%%%%%%%%%%%%%%%%%%%%%%%%%%%%%%%%%%%%%%%%%%%%%%%%%%%%%%%%%
\acknowledgments
%%%%%%%%%%%%%%%%%%%%%%%%%%%%%%%%%%%%%%%%%%%%%%%%%%%%%%%%%%%%%%%

We thank Y.~P.~Jing, Jounghun Lee, Yasushi Suto, Masahiro Takada,
Naohisa Inada, and Andrey Kravtsov for discussions.
MO is supported in part by JSPS through JSPS Research Fellowship for
Young Scientists. CRK is supported by NASA through Hubble Fellowship
grant HST-HF-01141.01-A from the Space Telescope Science Institute,
which is operated by the Association of Universities for Research
in Astronomy, Inc., under NASA contract NAS5-26555.

%%%%%%%%%%%%%%%%%%%%%%%%%%%%%%%%%%%%%%%%%%%%%%%%%%%%%%%%%%%%%%%%%%%%%%%

\end{document}